\documentclass[12pt]{iopart}
\usepackage{multirow}
\usepackage{iopams}  
\usepackage{graphicx}
\usepackage{setstack}
\usepackage{longtable,lscape,booktabs}
\usepackage{array}
\usepackage{tabularx}
\usepackage{color}

\begin{document}

\title{Multilayered crystals of macroions under slit-confinement}

\author{E C O\u guz, R Messina and H L\"owen}

\address{Institut f\"ur Theoretische Physik II: Weiche Materie,
Heinrich-Heine-Universit\"at D\"usseldorf, 
Universit\"atsstra\ss e 1, D-40225 D\"usseldorf, Germany}

\ead{ecoguz@thphy.uni-duesseldorf.de}
\ead{messina@thphy.uni-duesseldorf.de}
\ead{hlowen@thphy.uni-duesseldorf.de}

\begin{abstract}
The crystalline ground state of macroions confined between two neutral parallel plates
in the presence of their homogeneously spread counterions is calculated by 
lattice-sum minimization of candidate phases involving up to six layers. 
For increasing  macroion density, a cascade
of solid-solid transitions is  found involving various multilayered crystals.
The cascade includes triangular monolayer and buckled bilayer as well as
rhombic, squared and triangular phase structures.
\end{abstract}




\maketitle

\section{Introduction}
Strong correlations in Coulomb systems lead to a variety of new effects which 
are absent for neutral particles, see e.g.\ Refs.\ \cite{Levin,rene_review_2009} for a review. 
Among those are  nonlinear screening effects \cite{LMH1,LMH2,HL_94,review}, 
charge inversion \cite{Levin3}, Coulomb criticality \cite{Fisher,Levin2},
like-charge attraction for multivalent ions \cite{Amico2,Messina_PRL_2000,DNA,Zacca} 
as well as exotic binary crystalline structures unknown for uncharged 
systems \cite{Leunissen_Nature_2005,Hynnien_PRL_2006}.

By using charged colloidal suspensions \cite{Murray_review} 
or dust particles in plasmas \cite{Morfill}, it is possible
to realize strongly asymmetric mixtures of oppositely charged particles. These systems
consist of mesoscopic highly charged "macroions" and microscopic
counterions with a low valency resulting in strong charge and size asymmetries.
Since the charges of the macroions are high, strong Coulomb correlations 
are typical for macroions.
Most of the physics can still be encaptured by viewing these systems as 
strongly asymmetric and strongly coupled 
electrolytes. In recent years, it was possible to confine
macroions in sheets between two parallel plates 
\cite{Fontecha_2008,Cohen,Murray_review,Klapp_Zeng,Lobaskin} 
and to observe the resulting lateral structure of the particles. 
The gross features can be understood in terms of an (effective) one-component system with
a Yukawa pair interaction \cite{Kramposthuber,triplet,Damico,Russ_PRE,Russ,Dobnikar}. 
In fact, the mono- and bilayer ground-state structures which were
obtained  from a Yukawa model \cite{Messina_PRL}
describe the experimentally found structures \cite{comparative}. For multilayers
beyond the bilayer regime, a rich variety of stable phases are found in experiments 
\cite{Palberg_PRL_1997,Manzano_2007,Fontecha_2007} as well as in simulations \cite{Fortini},
which are all theoretically confirmed for a Yukawa system between two neutral walls \cite{Erdal_EPL}.
This motivates a study about the influence of the wall-particle interaction
on the phase behaviour of multilayered crystalline sheets in 
slit-like confinement \cite{Klapp}.

In this paper, we consider a model for macroions confined between two parallel neutral 
walls \footnote{Different from \cite{Goldoni} we include here a neutralizing background 
of counterions}. 
There is a direct Coulomb interaction between the point-like particles. 
The total system is charge-neutral and the counterions are kept at high temperature 
and are homogeneously spread between the plates resulting in an attraction acting 
on the macroions towards the middle of the plates. The system is realized for highly 
charged colloidal particles or dust particles in plasmas. Some early theoretical and simulational 
investigations on clusters of artificial atoms \cite{Sergio,Cornelissens,Reichhardt} and 
dusty plasmas \cite{Totsuji_PRL,Morfill} as well as one-component plasmas 
\cite{Rahman,Totsuji_Barrat,Schiffer_PRL}, including all the parabolic potentials 
acting as confinement, reveals the existence of multilayers.  
We therefore include the regime beyond bilayers in our discussion. 
Lattice sum minimizations among a broad set of candidate structures are used to 
determine the structure which minimizes the potential energy per particle. 
For increasing  macroion density, we find a cascade 
of solid-solid transitions which includes triangular monolayer, buckled bilayer and 
squared, rhombic and triangular bi-, tri-, tetra-, penta- and hexalayers
\footnote{For colloid-polymer films, see \cite{Ren}}.
Comparing the results to those involving a Yukawa interaction \cite{Erdal_EPL}, we show
that the topology of the phase diagram depends crucially on the particle-wall interaction.
In fact, some complicated tetralayered structures which were found stable for the 
confined Yukawa model are unstable in the present model. The strong correlation between phase
behaviour and wall-particle interactions suggests to tailor new crystalline
structures (e.g.\ with desired filtering properties \cite{Goedel})
by a suitable surface treatment of the plates.

The paper is organized as follows: the model is introduced in section II.
After discussing the structure of different crystalline multilayers, results
for the cascade of solid-solid transition are presented in section III.
Finally we conclude in section IV.

\section{The Model}

\begin{figure}[h!]
  \center
  \includegraphics[width=10cm]{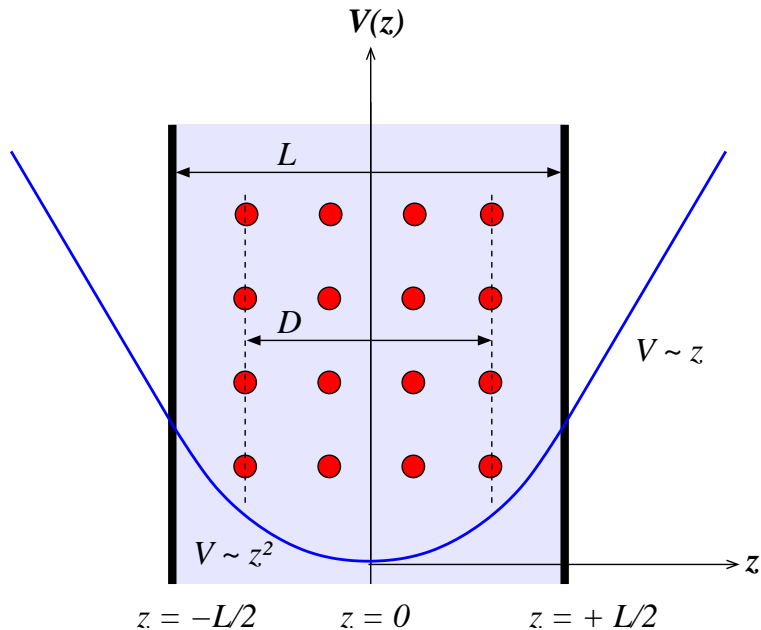}
  \caption{A schematic illustration of the model. 
    The ions (e.g., charged colloids) are represented by filled circles. 
    The counterions are smeared out 
    between the two hard walls located at $z=\pm L/2$.  
    This charge distribution generates a quadratic potential $V(z) \sim z^2$ 
    in between as shown.  
    The separation between outermost layers (dashed lines) is denoted by $D$.
    }
  \label{fig1}
\end{figure}

We consider $N$ classical point-like particles  
of charge $q$ ({\it macroions}) interacting via the unscreened Coulomb pair potential
%
%
\begin{equation}
  \label{coulomb}
  V(r)= \frac{q^2}{\epsilon r} ,
\end{equation}
%
%
where $r$ denotes the interparticle distance and $\epsilon$ the (relative) dielectric 
constant of surrounding medium. 
The system is confined between two parallel hard walls  of area $A$ and
separation $L$, see figure  \ref{fig1}. 
The global charge neutrality of the system is ensured by counterions. 
The latter are taken into account by an homogeneous neutralizing
background that is smeared out over the whole slit.  
We mention that we neglect the discrete nature of the counterions in this approach,
as well as any local ion-counterion coupling.
%
%
%

As a consequence of Gauss law, the electric field $E_b$
(stemming from the neutralizing background) is linear in $z$
inside the slit and constant outside the slit. 
More specifically, we have
%
%
\begin{equation} 
 \label{E_bg}
  E_b(z) = 
  \cases{
    - \frac{4\pi}{\epsilon} \frac{Nq}{A} \frac{z}{L} & for \quad $-L/2 \leq z \leq + L/2$, \\
    - \frac{2\pi}{\epsilon} \frac{Nq}{A} \frac{z}{|z|}  & else. 
  }  
\end{equation}
%
%
We thereby implicitly neglect image charge effects \cite{rene_jcp_2002}, 
meaning that we assume that there is no dielectric contrast at the interfaces (at $z = \pm L/2$).  
The resulting electrostatic potential $\Phi_b$, verifying the matching condition
at $z=\pm L/2$, then reads
%
\begin{equation} 
 \label{phi_bg}
  \Phi_b(z)  = 
  \cases{
  \frac{2\pi \eta q}{\epsilon L^3}  {z}^2 & for \quad $-L/2 \leq z \leq + L/2$, \\
  \frac{2\pi \eta q} {\epsilon L^2}  |z| - \frac{\pi \eta q}{2\epsilon L} & else, 
  } 
\end{equation}
%
%
where the reduced density 
%
\begin{equation} 
 \label{reduced-density}
  \eta \equiv \frac{N}{A} L^2
\end{equation}
%
was introduced. Hence, the potential of interaction $V_b(r)$ between a 
macroion and the counterion background is merely given by  
%
\begin{equation} 
 \label{V_macroion_bg}
  V_b (z) = q \Phi_b (z).
\end{equation}
%

We are now in a position to write the total potential energy per particle $u$ as
\footnote{
To remedy the divergence occurring with the first term of (\ref{u}),
a {\it two-dimensional} neutralizing background is introduced in the Lekner (or equivalently Ewald) sum. 
This neutralizing background (implicitly present in the Lekner and/or Ewald sum) 
has to be distinguished from the one that 
we use to model the counterions, which is smeared out over the whole {\it volume}
of the slit.
}
%
\begin{equation} 
  \label{u}
  u = \frac{1}{2N} \sum_{i=1}^{N} \sum_{j=1}^{N} V (r_{ij}) + 
 \frac{1}{N} \sum_{i=1}^{N} V_b (z_i).  
\end{equation}
%
In its appropriate rescaled form, $u$ reads (within the slit)
%
\begin{equation} 
 \label{u_rescaled}
 u \frac{\epsilon L}{q^2} 
 = 
 \frac{1}{2N} \sum_{i=1}^{N} \sum_{j=1}^{N} \frac{1}{r^*_{ij}}  + 
 \frac{1}{N} \sum_{i=1}^{N} 2\pi \eta   {z^*_i}^2,  
\end{equation}
%
with $r_{ij}^* \equiv r_{ij}/L$ and $z^*_i=z_i/L$, showing that at prescribed confinement width $L$
the energy of the system depends only on $\eta$. 
Consequently the phase diagram at zero temperature is given as a function of $\eta$.

At each given density $\eta$, we have performed lattice sum minimizations 
for a broad set of candidates of crystalline lattices. 
In order to handle the long ranged Coulomb potential, 
we have used the Lekner summation method \cite{Lekner} for three-dimensional 
systems with two-dimensional periodicity \cite{Brodka}, see also \cite{Mazars}.
More explicitly, we consider in this work three-dimensional crystals 
with two-dimensional periodicity in $x$- and $y$-direction 
whose primitive cell is a parallelepiped containing $n$ particles. 
This parallelepiped is spanned by the three lattice vectors ${\bf a}=a(1,0,0)$, 
${\bf b}=a\gamma(\cos\theta,\sin\theta,0)$ and ${\bf c}=D(0,0,1)$, where $\gamma$ is the aspect ratio 
($\gamma = |{\bf b}|/|{\bf a}| = b/a$) and $\theta$ is the angle 
between ${\bf a}$ and ${\bf b}$. Furthermore, the $n$ particles are distributed, 
not necessarily evenly, on $m$ layers in the $z$-direction  
such that $c=|{\bf c}|$ corresponds to the distance between outermost layers (see also figure \ref{fig1}).
Hereby we restrict ourselves to layered situations with an up-down 
inversion symmetry in the averaged occupancy reflecting the up-down symmetry
of the confining slit. Under this sole restriction, we consider possible candidates 
with $n=1,\cdots,8$ and $m=1,\cdots,6$ up to symmetric six-layer structures with 
a basis of up to 8 particles. Furthermore, we also examine the stability of several asymmetric  
buckling phases, as predicted in \cite{Chou_Nelson}. For given $\eta$, the total potential energy 
per particle is minimized with respect to the particle coordinates of the basis 
and the cell geometry ($\gamma$ and $\theta$). The resulting stability 
phase diagrams are shown and discussed in the following sections.

\section{Mono- and bilayer phase behavior}

\subsection{Phase diagram}

An increase of  $\eta$ within the mono- and bilayer regime reveals the existence of five stable 
crystalline mono- and bilayers: $1\Delta$ (triangular), $3\Delta$ (staggered triangular), 
$2\square$ (square), $2R$ (rhombic) and $2\Delta$ (staggered triangular). 
The integers indicate the number of layers. For increasing $\eta$, 
the stability cascade therefore reads: 
%
\begin{equation}
  \label{seq_bi}
  1\Delta \to 3\Delta \to 2\square \to 2R \to 2\Delta .
\end{equation}
%
%

\begin{figure}[h!]
  \center
  \includegraphics[width=11cm]{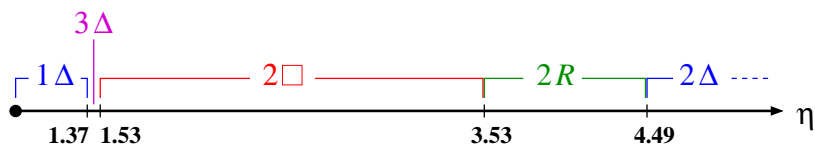}
  \caption{Stability phase diagram of crystalline mono- and bilayers. 
    The five stable phases $1\Delta$, $3\Delta$, $2\square$, $2R$ and $2\Delta$ correspond to 
    Wigner crystals, found in earlier investigations (see text for details). 
    Note that the monolayer-trilayer transition occurs at $\eta \approx 1.37$.}
  \label{phase_diag1}
\end{figure}

Most of these phases, corresponding to Wigner crystals predicted in earlier 
theoretical investigations \cite{Goldoni,Messina_PRL}, 
are also found in experiments on charged colloidal suspensions \cite{Pansu1984,Neser1997} 
as well as in Monte Carlo simulations of confined hard spheres \cite{Schmidt_PRL}. 
The detailed phase diagram is reported in figure \ref{phase_diag1}. 
%
%
%

We emphasize that the $3\Delta$ phase (staggered in an $ABC$ manner, see also table \ref{tab}) 
intervenes between $1\Delta$ and $2\square$ rather than a buckled phase which is present 
in a situation where the external potential has a vanishing curvature at the origin.

At small reduced densities $\eta$, particles tend to stay in the 
potential minimum (cf.\ figure \ref{fig1}) created by the counterion background. 
This  is precisely the origin of the stability of monolayered Wigner crystals, 
which never occurs in purely unscreened Coulomb systems.
\footnote{Indeed, we found that a rectangular bilayer with size ratio $\gamma = \sqrt{3}$, 
proposed as a stable structure for very small $\eta$ in \cite{Goldoni}, is always energetically 
beaten by a buckled ($2B$) bilayered phase. Seen from the top, this structure corresponds to the 
triangular lattice.}  
The triangular monolayer $1\Delta$ is stable up to $\eta = 1.37$. 
At larger densities the mutual repulsive interparticle interactions, first term in 
equation (\ref{u_rescaled}), 
dominates the competition between the interparticle (macroion-macroion) repulsion 
and particle-background (macroion-counterion) attraction.   

%
\begin{figure}
  \center
  \includegraphics[width=11cm]{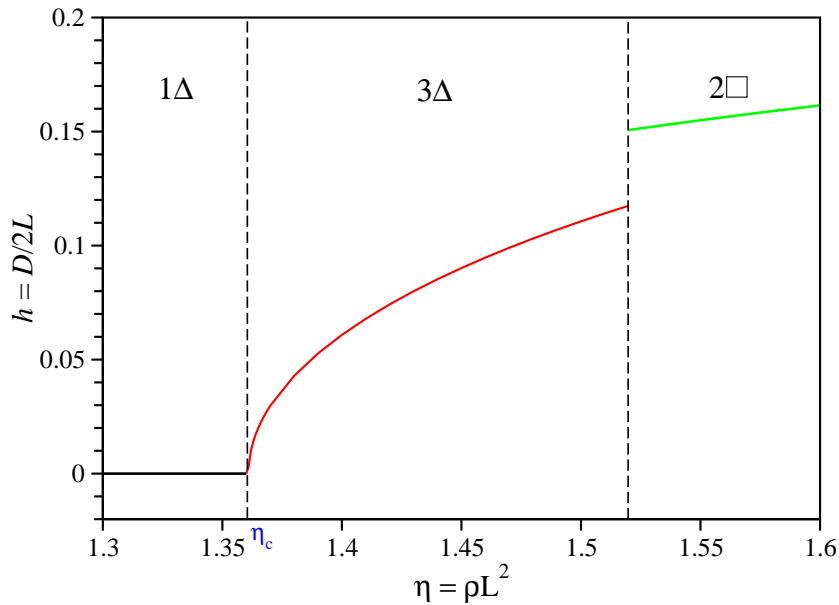}
  \caption{Order parameter $h$ in the transition regime $1\Delta$ to $2\square$ via $3\Delta$. 
    The monolayer $1\Delta$ buckles at a critical density $\eta_c \sim 1.360901$ to a trilayer.}
  \label{fig3}
\end{figure}
%

The structure with triangular base shape $3\Delta$ appears as the first stable multilayer 
(see figure \ref{fig3}), interpolating between $1\Delta$ and $2\square$.
The associated order parameter, namely the reduced separation
%
\begin{equation} 
 \label{h-layer}
  h \equiv \frac{D}{2L}
\end{equation}
%
between the mid-plane and the outer macroion layer (see also figure \ref{fig1}), 
is continuous at the transition $1\Delta \to 3\Delta$ but 
discontinuous across the $3\Delta \to 2\square$ transition, see figure \ref{fig3} and \cite{Schmidt_PRL}.

\begin{figure}
  \center
  \includegraphics[width=11cm]{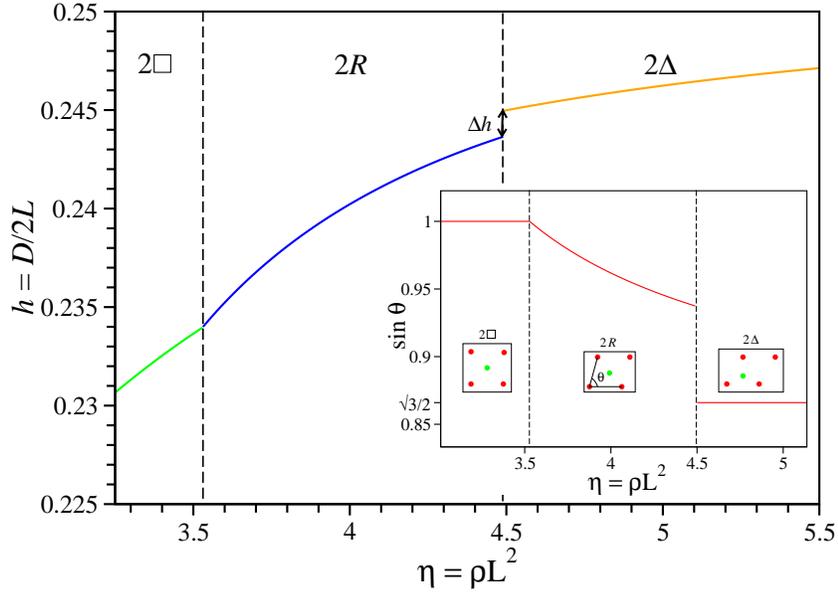}
  \caption{Order parameter $h$ in the transition regime $2\square$ to $2\Delta$ via $2R$.
    The discontinuity $\Delta h$ in the developing of the layer-layer separation by the transition 
    $2R \to 2\Delta$ is also shown for clarity. In the inset one can regard how $\theta$ changes 
    in the same regime. Corresponding structures are also sketched in the inset.  
    Different colors indicate different layers. }
  \label{fig4}
\end{figure}

By further increase of $\eta$, one recovers the 
rhombic phase $2R$, which is continuously achievable from the square phase $2\square$ by changing $\theta$, 
as indicated in the inset of figure \ref{fig4}. 
The two geometrical order parameters $h$ and $\sin \theta$, see figure \ref{fig4}, 
indicate thereby  a continuous transition for $2B \to 2\square$.
On the other hand, at larger values of $\eta$, the transition 
$2R \to 2\Delta$ is of first order as signaled by the jumps of
the two geometrical order parameters $h$ and $\sin \theta$,
see figure \ref{fig4}.
The staggered triangular phase $2\Delta$ corresponds to the 
ultimate stable structure in the high density regime of bilayers.
%

%
%
%

\subsection{From monolayer to trilayer - An analytic approach}

\begin{figure}
  \center
  \includegraphics[width=11cm]{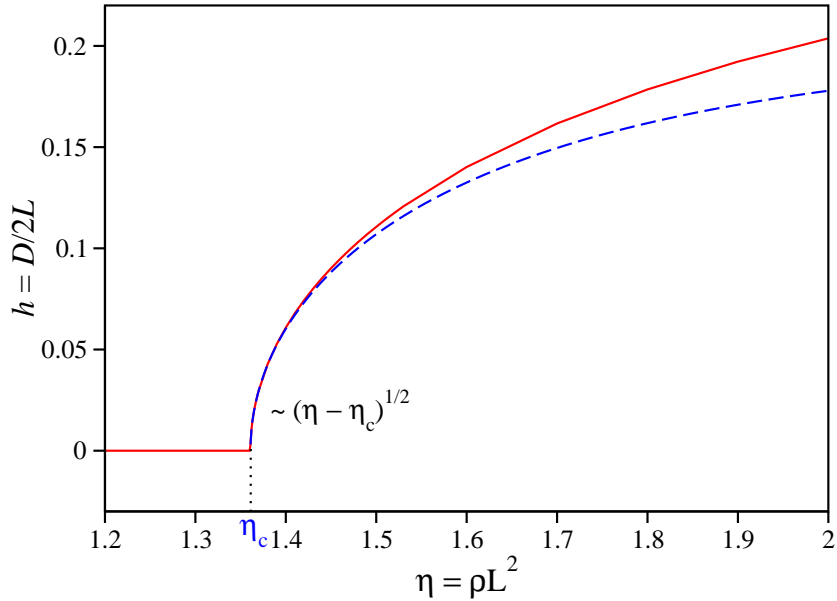}
  \caption{Plot of equation equation (\ref{eq.h2_critic}) (dashed line) and numerical calculations  
    for finite $h(\eta)$ (solid line) based on full lattice sum minimization 
    near the monolayer-trilayer $1\Delta \to 3\Delta$ transition. }
  \label{fig5}
\end{figure}

We now would like to address the transition $1\Delta \to 3\Delta$ analytically. 
To do so, we apply a Taylor expansion  to $u(h)$ around $h=D/2L=0$, see the Appendix for details. 
The resulting asymptotic expression for small interlayer distances $h$ reads 
%
\begin{equation}
  \frac{u(h)}{ q^2 / \epsilon L} = 
  B_0 \sqrt \eta + B_1 \eta^{3/2} h^2 + B_2 \eta^{5/2} h^4 + \frac{4}{3}\pi \eta h^2 .
  \label{pot_taylored}
\end{equation}
%
with
%
\begin{equation}
  B_0 = -1.960516\dots, \quad  B_1 = - 3.590668\dots, \quad  B_2 = 4.968827\dots.
  \label{pot_taylored_coeff}
\end{equation}
%
%
The profile of the reduced half layer-layer distance $h(\eta)$ is obtained upon minimizing 
$u$ with respect to $h$, i.e. $\partial{u}/\partial h = 0$, 
leading to
%
\begin{equation}
  \label{h_eta}
  h^2 (\eta) = - \frac{B_1\sqrt{\eta} + \frac{4}{3}\pi}{2B_2{\eta}^{3/2}} .
\end{equation}
%
%
It is now a simple matter to obtain the reduced density $\eta_c$
at which the monolayer-trilayer transition ($1\Delta \to 3\Delta$) takes place.
The mathematical condition is thereby $h(\eta=\eta_c)=0$ yielding
%
\begin{equation}
  \label{eta_crit}
  \sqrt{\eta_c} = -\frac{4\pi}{3B_1} \Rightarrow \eta_c = 1.360901\dots,
\end{equation}
%
%
which is in quantitative agreement with the lattice sum minimization 
results from  previous section, see figure \ref{fig5}. 

By inserting the expression (\ref{eta_crit}) of $\eta_c$ in (\ref{h_eta}) 
one obtains 
%
\begin{equation}
  \label{eq.h2_critic}
  h^2(\eta) = -\frac{B_1}{2B_2} \frac{\eta - \eta_c}{{\eta}^2 + {\eta}^{3/2}\sqrt{\eta_c}}.
\end{equation}
%
Noticing that the last denominator in equation (\ref{eq.h2_critic}) 
can be approximated  (valid in the relevant limit $\eta \to \eta_c^+$) by $2\eta^2$,
we obtain a square-root singularity:
%
\begin{equation}
  \label{h_trans_crit}
  \lim_{\eta \to \eta_c^+ } h(\eta) =
  \sqrt{-\frac{B_1}{4B_2 \eta_c^2}} (\eta - \eta_c)^{1/2}
  \sim (\eta - \eta_c)^{1/2}.
\end{equation}
%
This theoretical prediction (\ref{eq.h2_critic}) is visualized in figure \ref{fig5}. 
%
%

\section{Multilayers}

The presence of the neutralizing background allows the formation of multilayers with $m \geq 3$ 
for large enough densities $\eta$, which is forbidden in the absence of a background 
\footnote{There is a simple and clear electrostatic argument to explain the exclusive 
  stability of bilayers for charges confined between (charged or uncharged) hard walls 
  without neutralizing {\it volume} background. One has to note that two equally charged walls 
  do {\it not} generate any electric field within the slit, and consequently do not alter 
  the stable structure obtained at any other surface charge (including neutral walls). 
  Hence, if one considers the special case of two walls corresponding to {\it two-dimensional} 
  neutralizing backgrounds where the ground-state is the $2\Delta$ bilayer, we deduce from 
  this that the  ground state structure is {\it always} a bilayer.}. 
The physical origin of the stability of multilayers in the present system at large $\eta$ is basically 
a balance between the mutual {\it unscreened} macroion-macroion repulsion and 
the attractive macroion-background interaction. 

\begin{figure}
  \center
  \includegraphics[width=16cm]{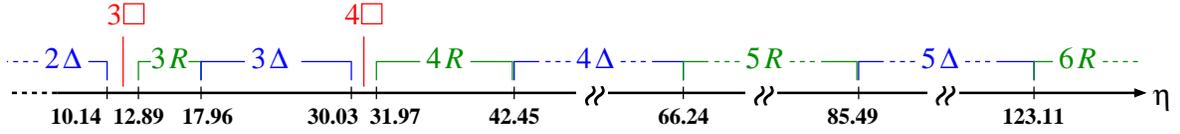}
  \caption{Stability phase diagram of crystalline multilayers in the presence of a neutralizing background. 
    $3\square$, $3R$, $3\Delta$, $4\square$, $4R$, $4\Delta$, $5R$, $5\Delta$ and $6R$ 
    are obtained as stable in the analyzed $\eta$-regime. The corresponding structures are given in table \ref{tab}. 
  }
  \label{fig6}
\end{figure}

We shall now analyze in detail the high density regime up to $\eta \approx 130$. 
Beyond the bilayer regime, that is limited by $2\Delta$, the cascade found here 
upon increasing $\eta$ reads: 
%
\begin{equation}
  \label{seq_multi}
  \fl
  \cdots 3\square \to 3R \to 3\Delta \to 4\square \to 4R \to 4\Delta \to 5R \to 5\Delta \to 6R \cdots ,
\end{equation}
%
%
where rhombic phases $3R$, $4R$, $5R$ and $6R$ have the stacking sequence $ABA$, $ABAB$, 
$ABABA$ and $ABABAB$ while the triangular phases $3\Delta$, $4\Delta$ and $5\Delta$ occur as 
$ABC$, $ABCA$ and $ABCAB$, respectively. 
More structural details are given in table \ref{tab}. 
The corresponding phase diagram is depicted in figure \ref{fig6}. 
%
%
%
\begin{landscape}
  \begin{center}
    \begin{indented}
    \item[]
      \begin{longtable}{@{}cccccccm{3.1cm}m{3.1cm}}
        \caption{\label{tab} Structural details and schematic illustration of the stable crystalline 
          multilayers. The layers are labeled as follows. The bottom one located at $z=-D/2$ 
          corresponds to first layer (labeled as $i=1$), and the labels of the successive layers 
          are incremented accordingly. For $m>3$, the separation between the two first layers is 
          characterized by $\delta D$ with $1/(m-1) \leq \delta < 0.5 $. The relative separation vector 
          between two particles of a primitive cell belonging to two layers $i$ and $j$ 
          is given by ${\bf d}_{ij}$. For six layers, the separation between the first and 
          the third layers is specified by $\lambda D$ with $2/5 \leq \lambda < 0.5$. 
          In the top views of $3\Delta$, $4\Delta$, $5\Delta$ and  $3R$, $4R$, $5R$, $6R$ 
          each basis shape (triangular or rhombic) is emphasized with white lines.
          The rhombic stripes of $3R$, $4R$, $5R$ and $6R$ are shown again in corresponding 
          perspective views, for clarity.
          Particles from different layers are identified by different colors.}
        \\[\abovecaptionskip]
        \br
        Phase & $\mathrm{{\bf b}/{\it a}}$ & ${\bf d}_{12}$ & ${\bf d}_{13}$ & ${\bf d}_{14}$ 
        & ${\bf d}_{15}$ & ${\bf d}_{16}$  & \hspace{0.75cm} top view & \hspace{0.1cm} side/persp. view 
        \\\br
        
        \vspace{-0.13cm}
        $3\square$ & $(0,1)$   &  $\displaystyle \mathrm{\frac{{\bf a} + {\bf b} + {\bf c}}{2}}$ 
        & ${\bf c}$  & -- & -- & -- 
        & \includegraphics[width=3.1cm]{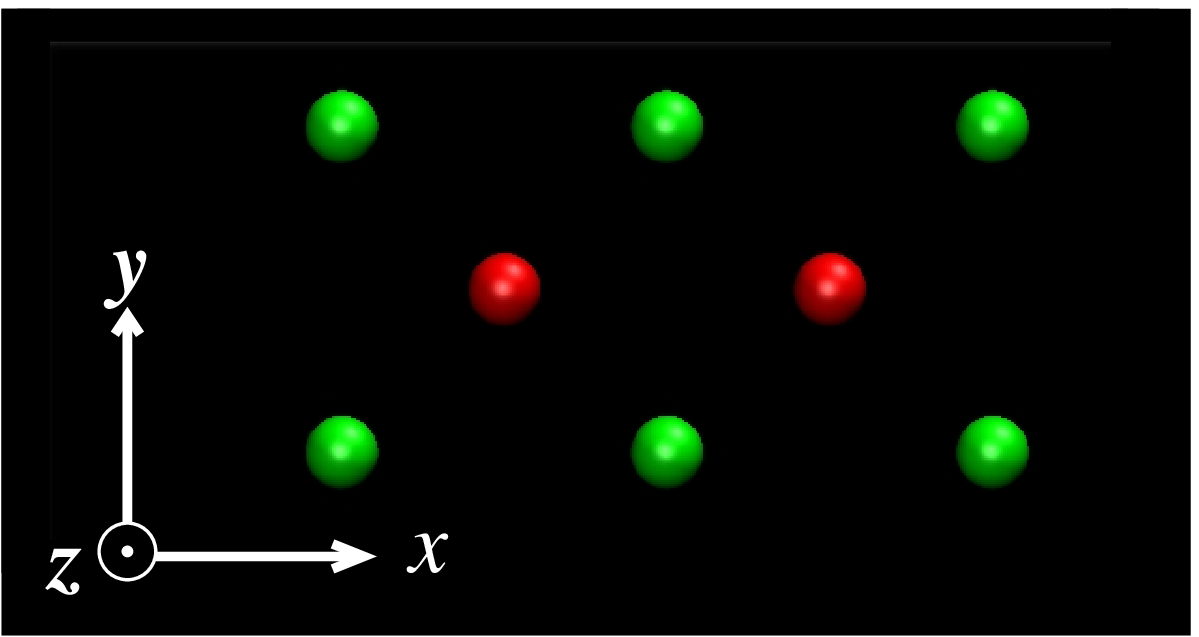} 
        & \includegraphics[width=3.1cm]{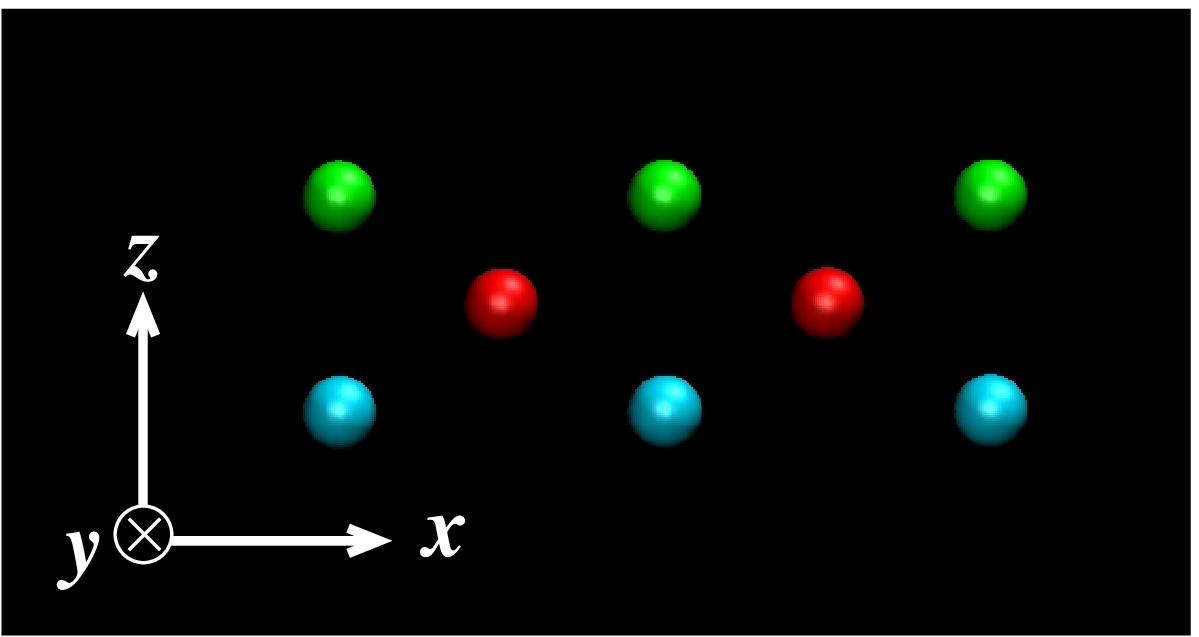} 
        \\
        \mr
        
        \vspace{-0.13cm}
        3R & $(\cos\theta,\sin\theta)$ & $\displaystyle \mathrm{\frac{{\bf a} + {\bf b} + {\bf c}}{2}}$ 
        & ${\bf c}$ & -- & -- & --
        & \includegraphics[width=3.1cm]{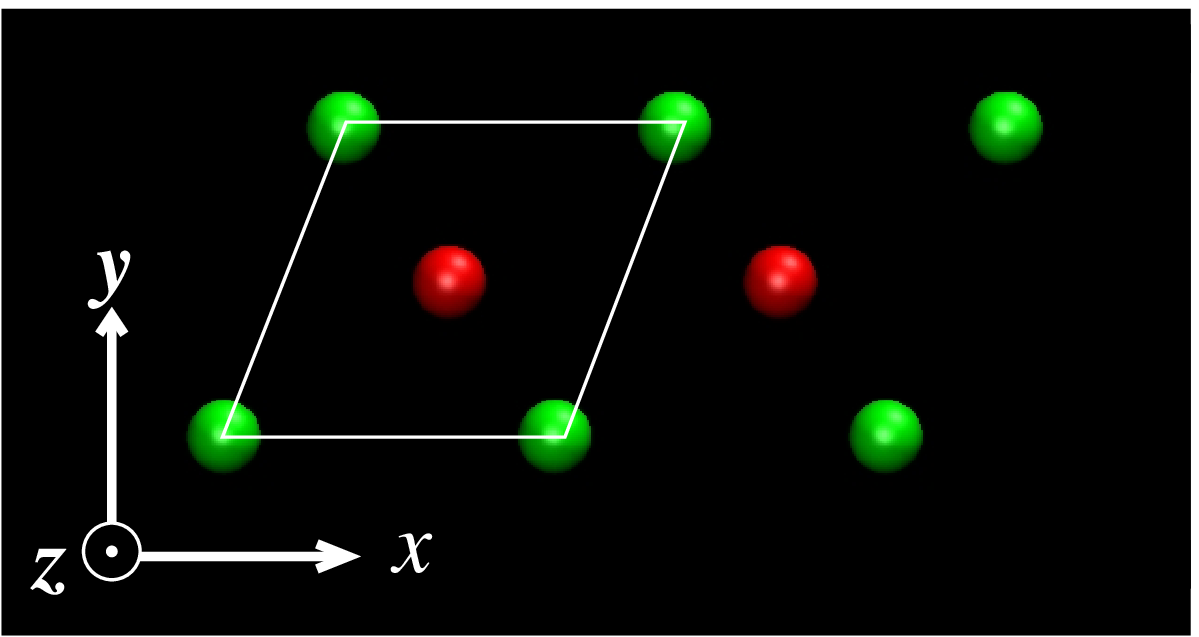} 
        & \includegraphics[width=3.1cm]{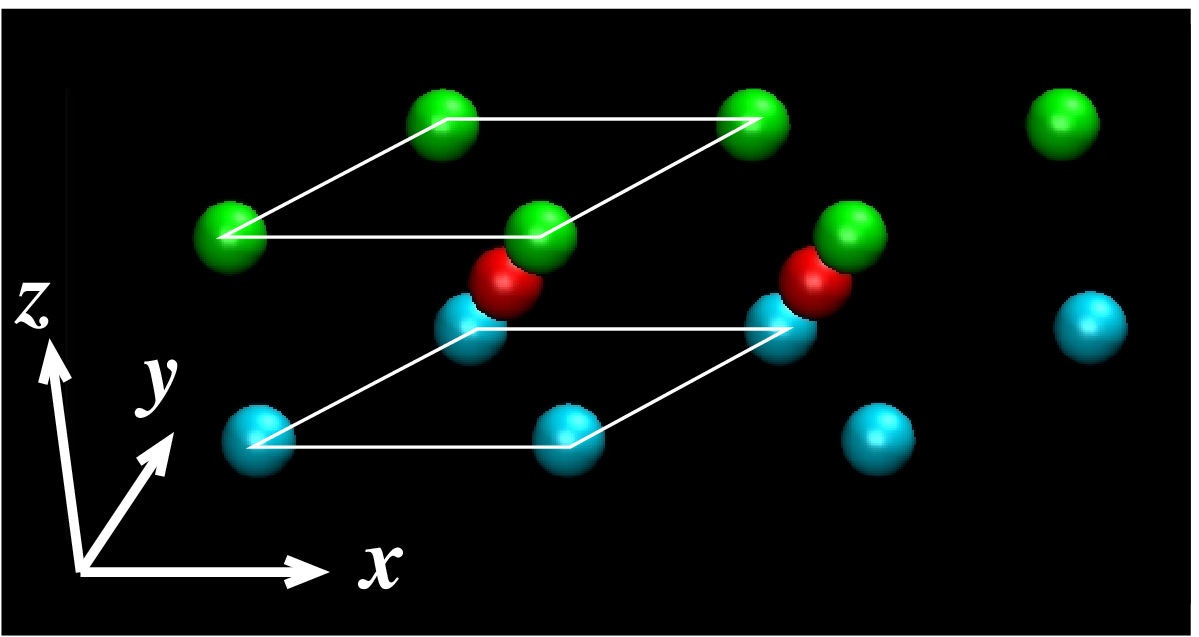}  
        \\\mr
        
        \vspace{-0.13cm}
        3$\Delta$  & $(1/2,\sqrt{3}/2)$  & $\displaystyle \mathrm{\frac{{\bf a}+{\bf b}}{3} + \frac{{\bf c}}{2}} $  
        & $\displaystyle \mathrm{\frac{2({\bf a}+{\bf b})}{3} + {\bf c}}$  &--&--&-- 
        & \includegraphics[width=3.1cm]{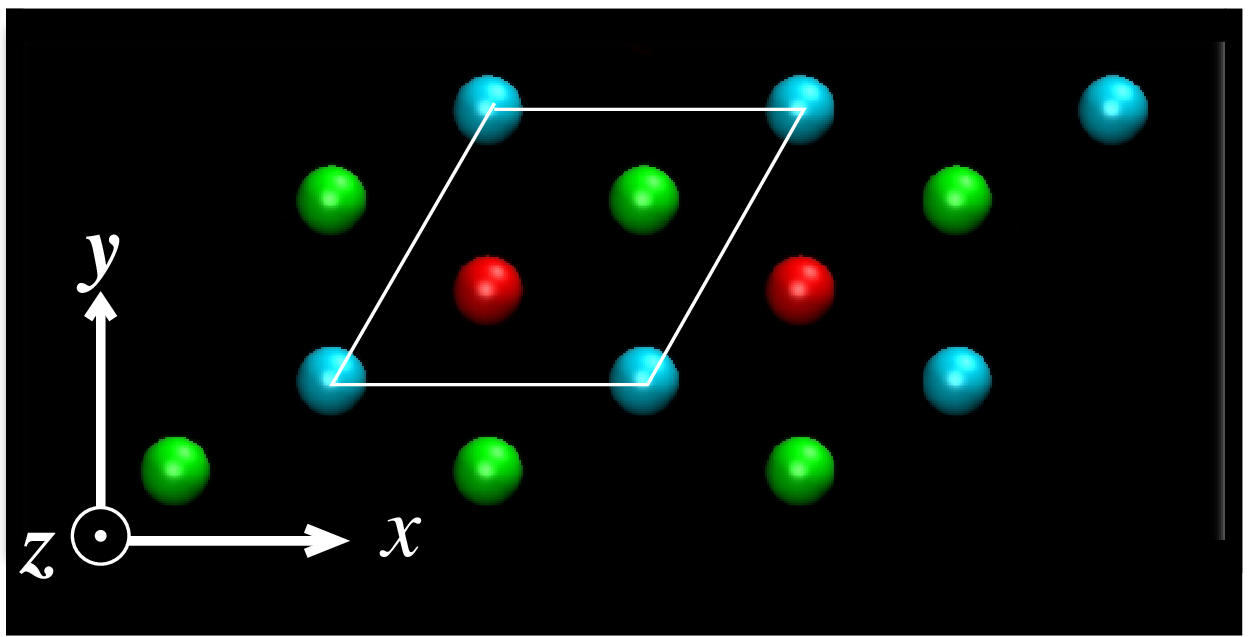} 
        & \includegraphics[width=3.1cm]{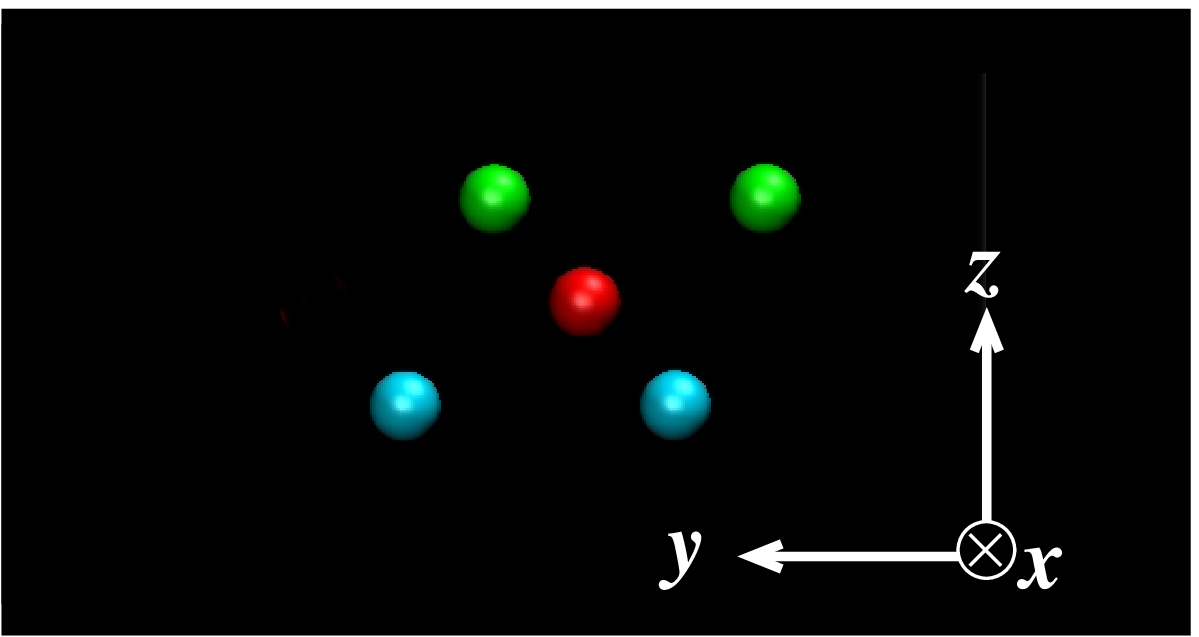}  
        \\\mr
        
        \vspace{-0.13cm}
        4$\square$ & $(0,1)$  & $\displaystyle \mathrm{\frac{{\bf a}+{\bf b}}{2} + {\bf c}\delta}$ 
        & ${\bf c}(1-\delta)$ & $\displaystyle \mathrm{\frac{{\bf a}+{\bf b}}{2} + {\bf c}}$ & -- &--
        & \includegraphics[width=3.1cm]{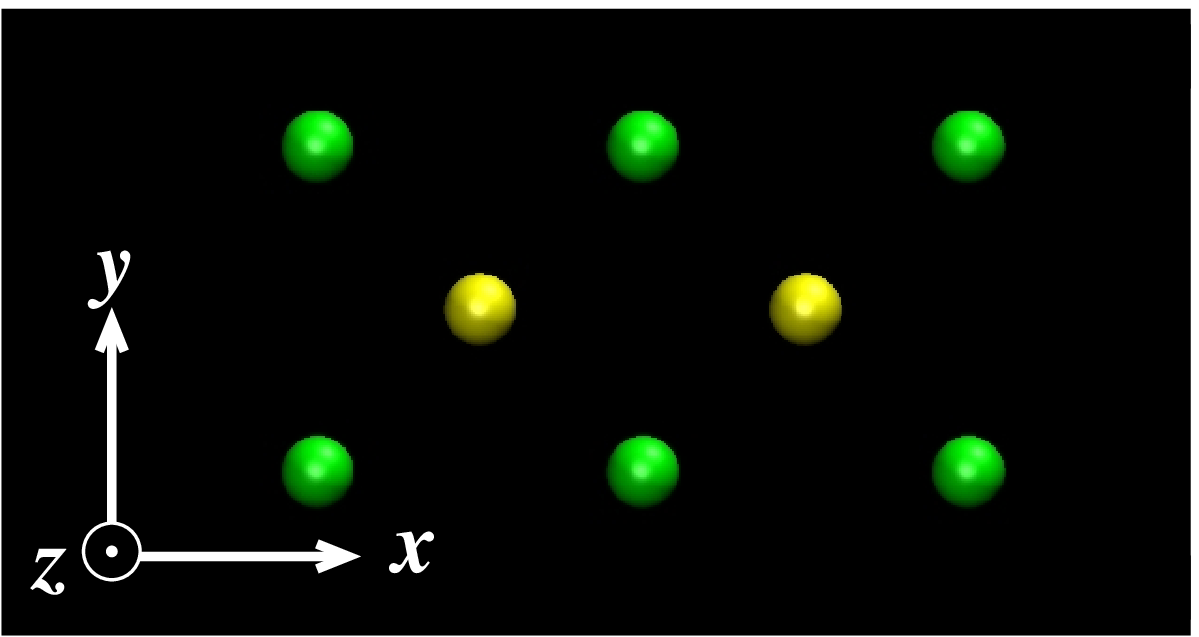} 
        & \includegraphics[width=3.1cm]{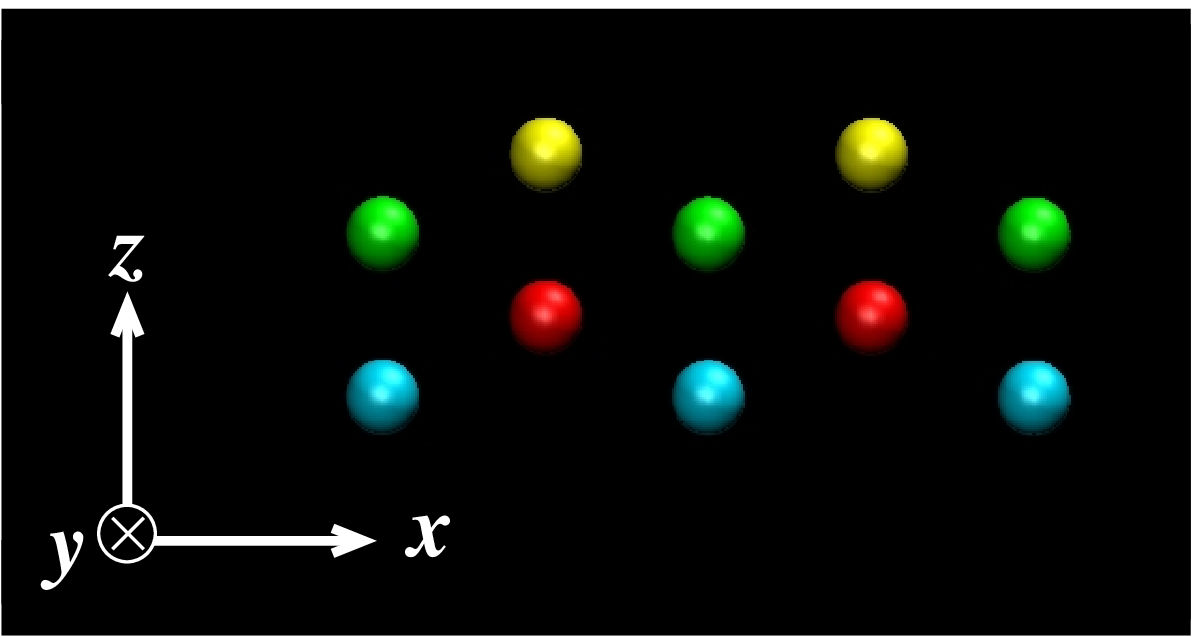}
        \\\mr
        
        \vspace{-0.13cm}
        4R & $(\cos\theta,\sin\theta)$ &  $\mathrm{\frac{1}{2}({\bf a}+{\bf b}) + {\bf c}\delta}$  
        & ${\bf c}(1-\delta)$ & $\mathrm{\frac{1}{2}({\bf a}+{\bf b}) + {\bf c}}$ &--& --
        &\includegraphics[width=3.1cm]{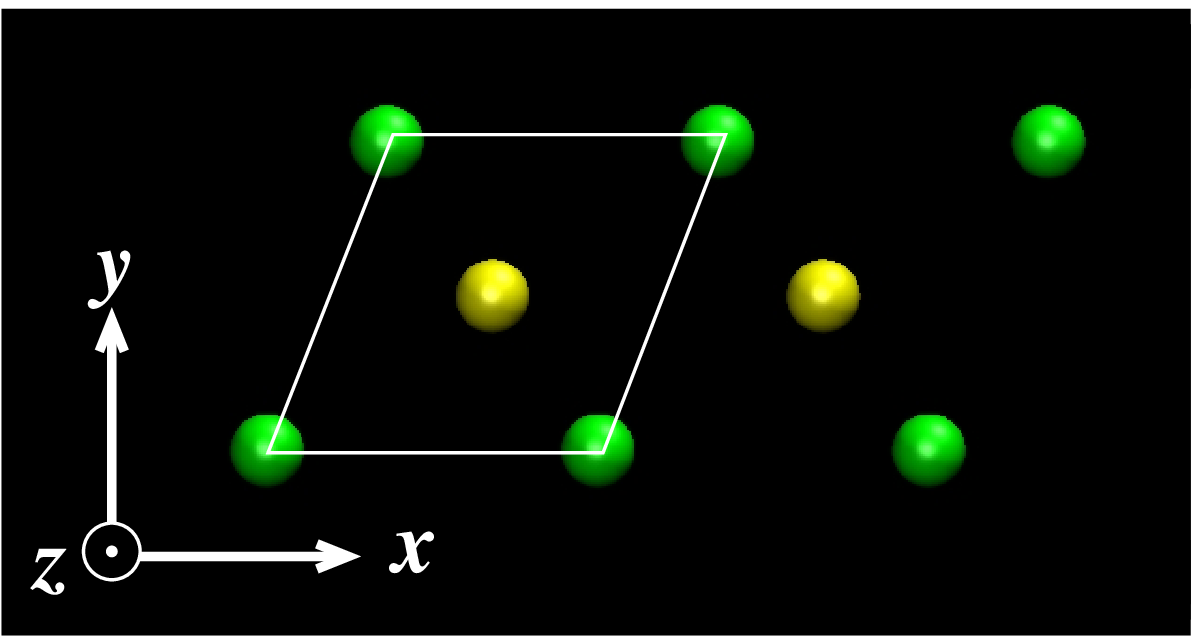}
        &\includegraphics[width=3.1cm]{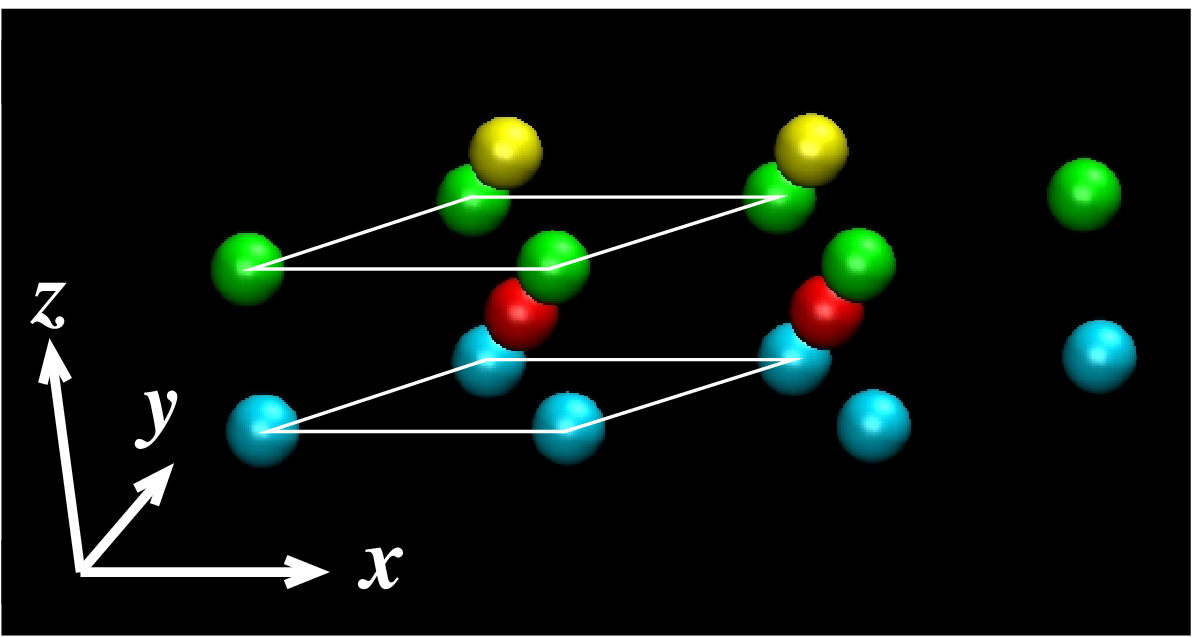}
        \\\mr

        \vspace{-0.13cm}
        4$\Delta$  & $(1/2,\sqrt{3}/2)$  &  $\displaystyle \mathrm{\frac{{\bf a}+{\bf b}}{3} + {\bf c}\delta}$  
        & $\displaystyle \mathrm{\frac{2({\bf a}+{\bf b})}{3} + {\bf c}(1-\delta)}$ & ${\bf c}$ &--&--
        & \includegraphics[width=3.1cm]{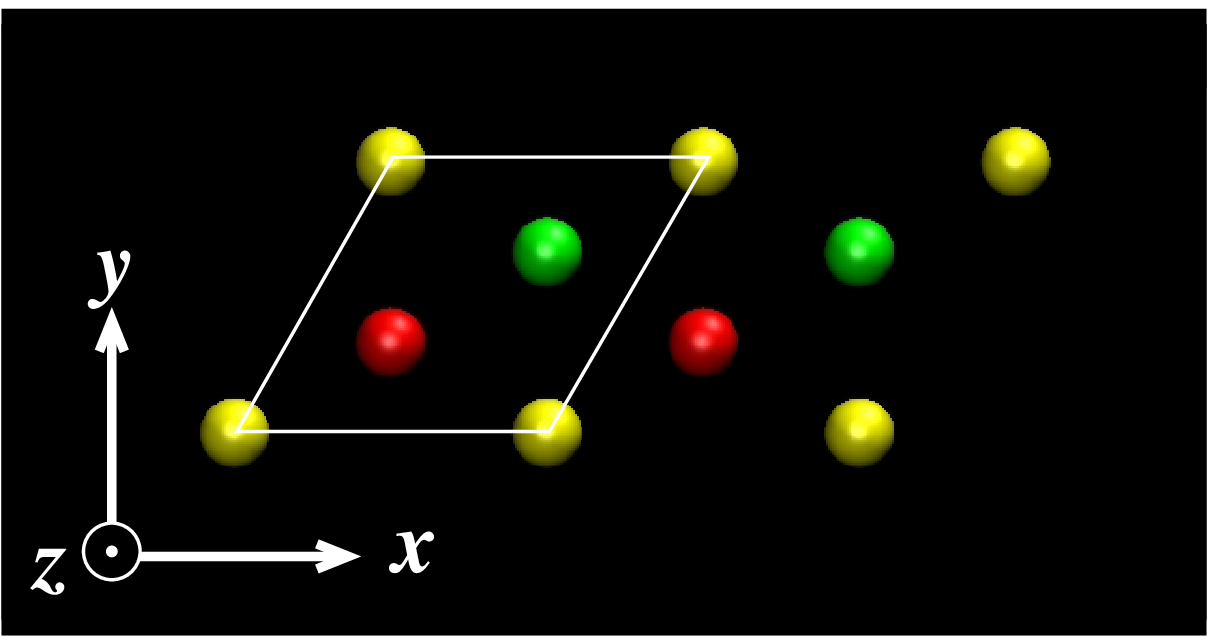}
        & \includegraphics[width=3.1cm]{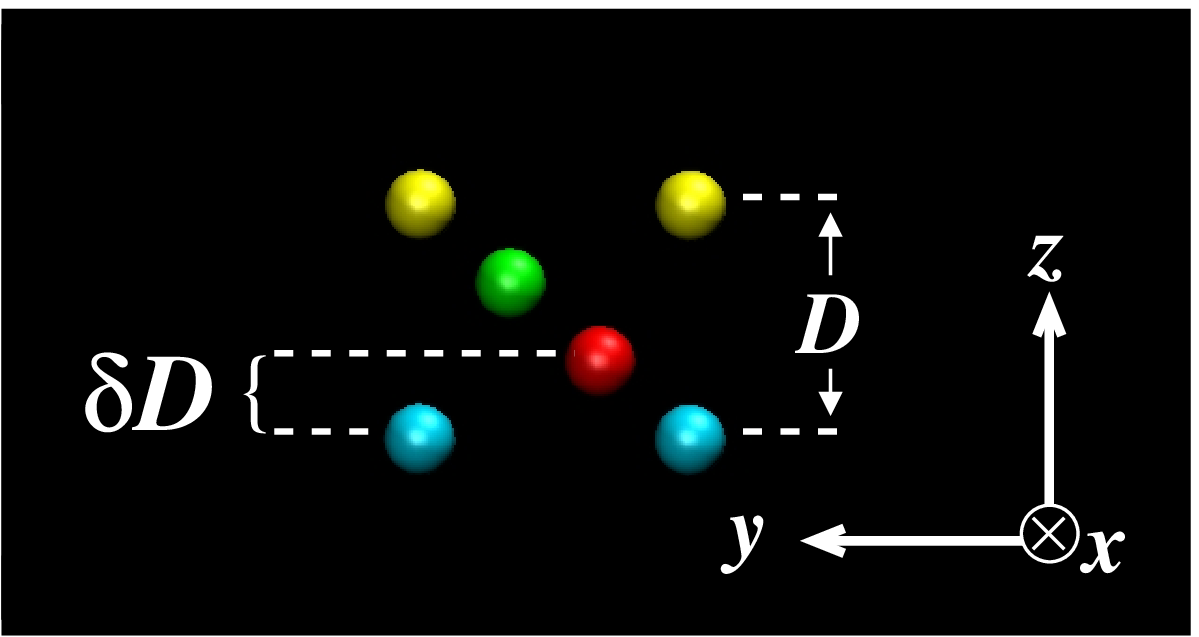}
        \\\mr

        \vspace{-0.13cm}
        5R & $(\cos\theta,\sin\theta)$ &  $\displaystyle \mathrm{\frac{{\bf a}+{\bf b}}{2} + {\bf c}\delta}$ 
        & $\displaystyle \mathrm{\frac{{\bf c}}{2}}$ 
        & $\displaystyle \mathrm{\frac{{\bf a}+{\bf b}}{2} + {\bf c}(1-\delta)}$ 
        & $\mathrm{{\bf c}}$ & --  
        & \includegraphics[width=3.1cm]{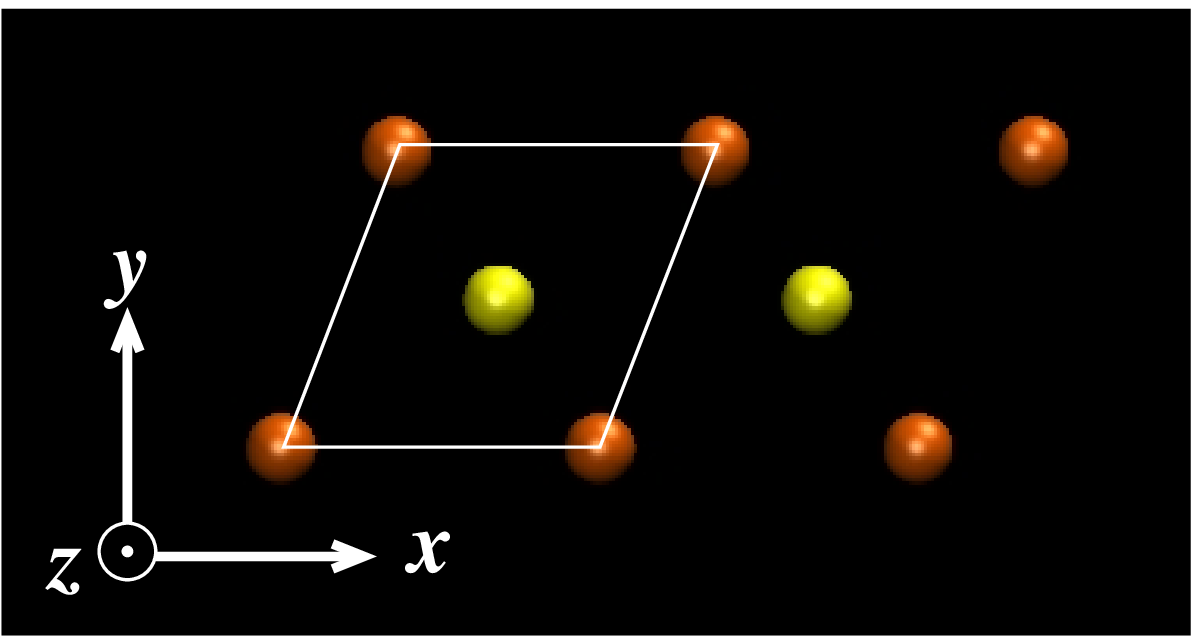} 
        & \includegraphics[width=3.1cm]{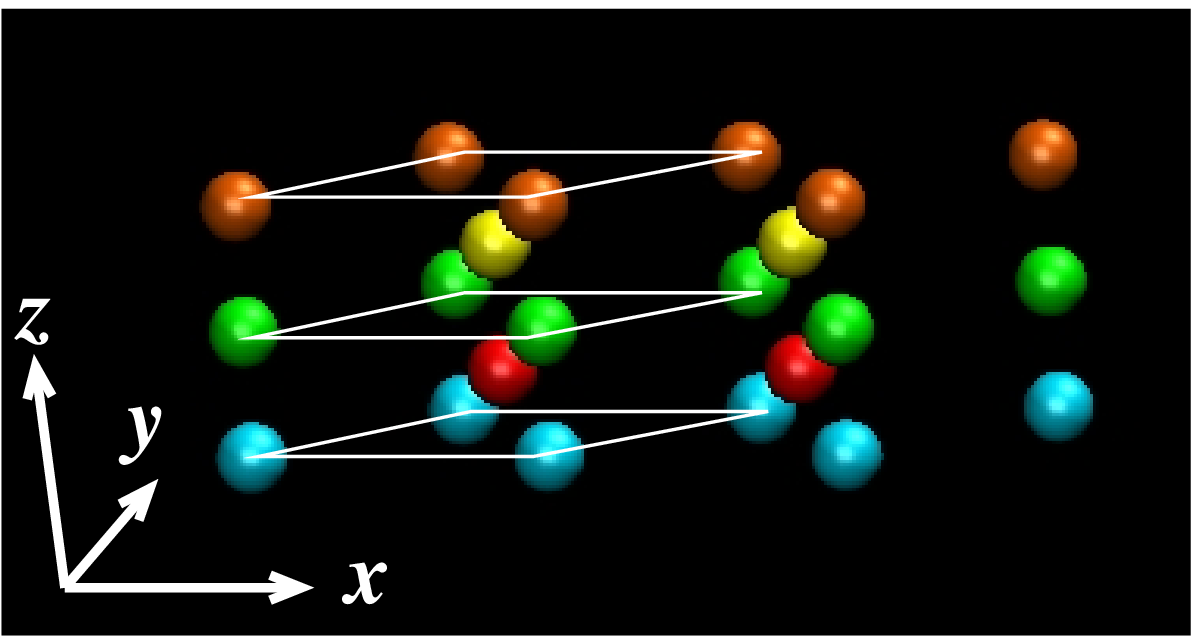} 
        \\\mr
        
        \vspace{-0.13cm}
        5$\Delta$ & $(1/2,\sqrt{3}/2)$ &  $\displaystyle \mathrm{\frac{{\bf a}+{\bf b}}{3} + {\bf c}\delta}$   
        & $\displaystyle \mathrm{\frac{2({\bf a}+{\bf b})}{3} + \frac{{\bf c}}{2}}$ & $\mathrm{{\bf c}(1-\delta)}$ 
        & $\displaystyle \mathrm{\frac{{\bf a}+{\bf b}}{3} + {\bf c}}$  & -- 
        & \includegraphics[width=3.1cm]{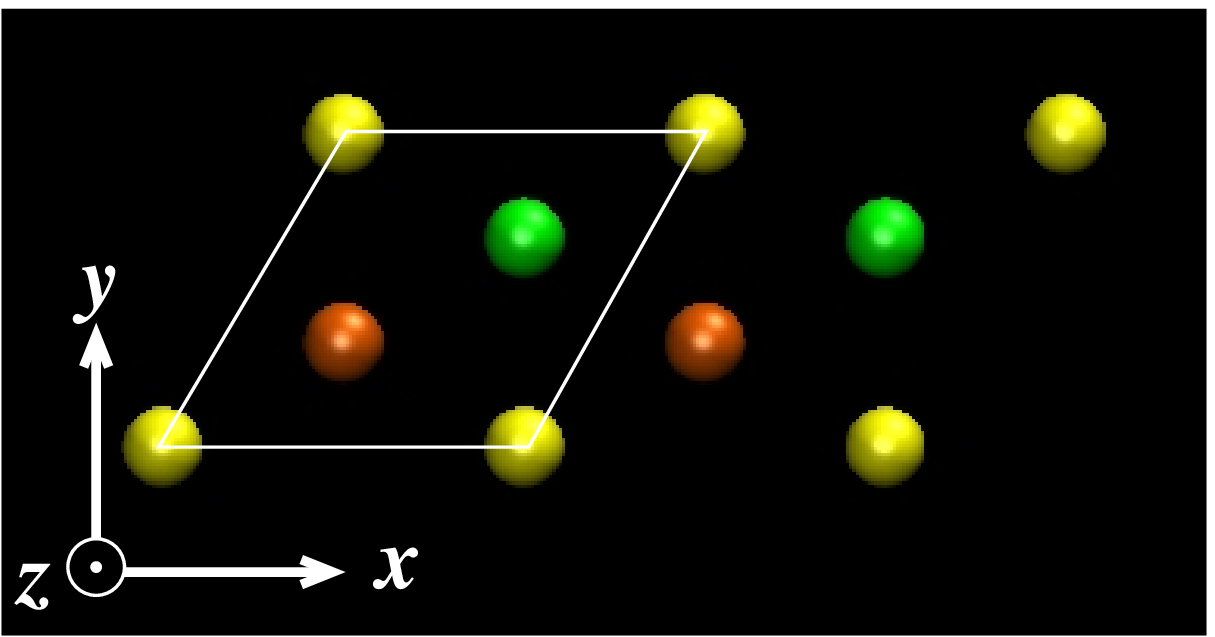} 
        & \includegraphics[width=3.1cm]{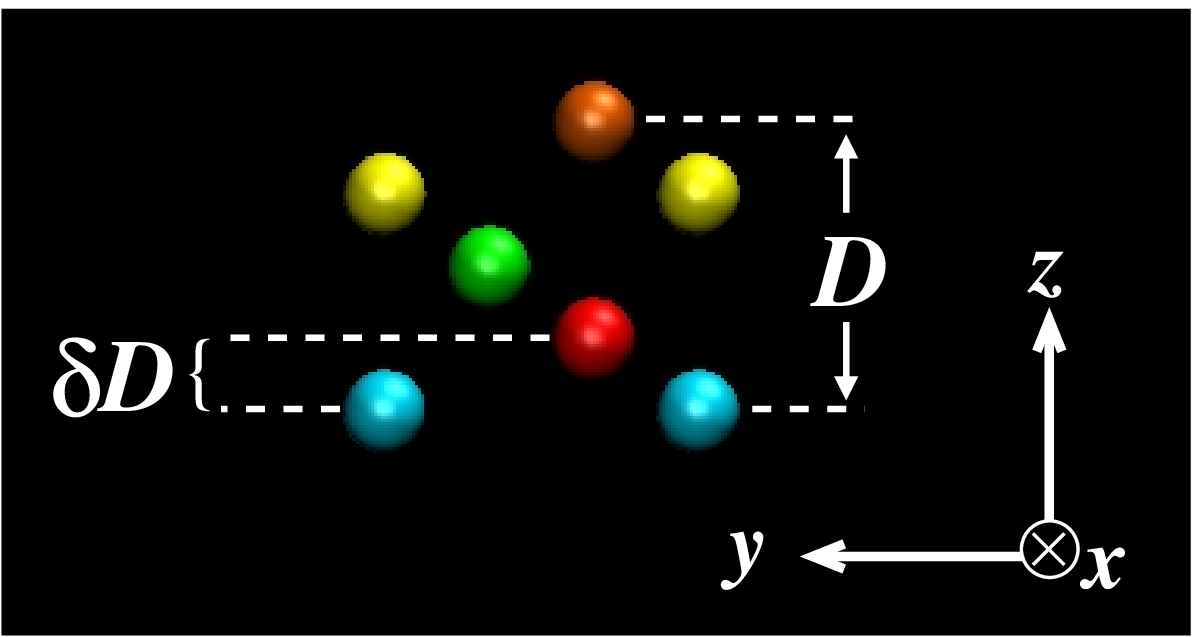} 
        \\\mr
        
        \vspace{-0.13cm}
        6R & $(\cos\theta,\sin\theta)$ & $\displaystyle \mathrm{\frac{{\bf a}+{\bf b}}{2} + {\bf c}\delta}$ 
        & $\mathrm{{\bf c}\lambda}$    & $\displaystyle \mathrm{\frac{{\bf a}+{\bf b}}{2} + {\bf c}(1-\lambda)}$ 
        & $\mathrm{{\bf c}(1-\delta)}$ & $\displaystyle \mathrm{\frac{{\bf a}+{\bf b}}{2} + {\bf c}}$  
        & \includegraphics[width=3.1cm]{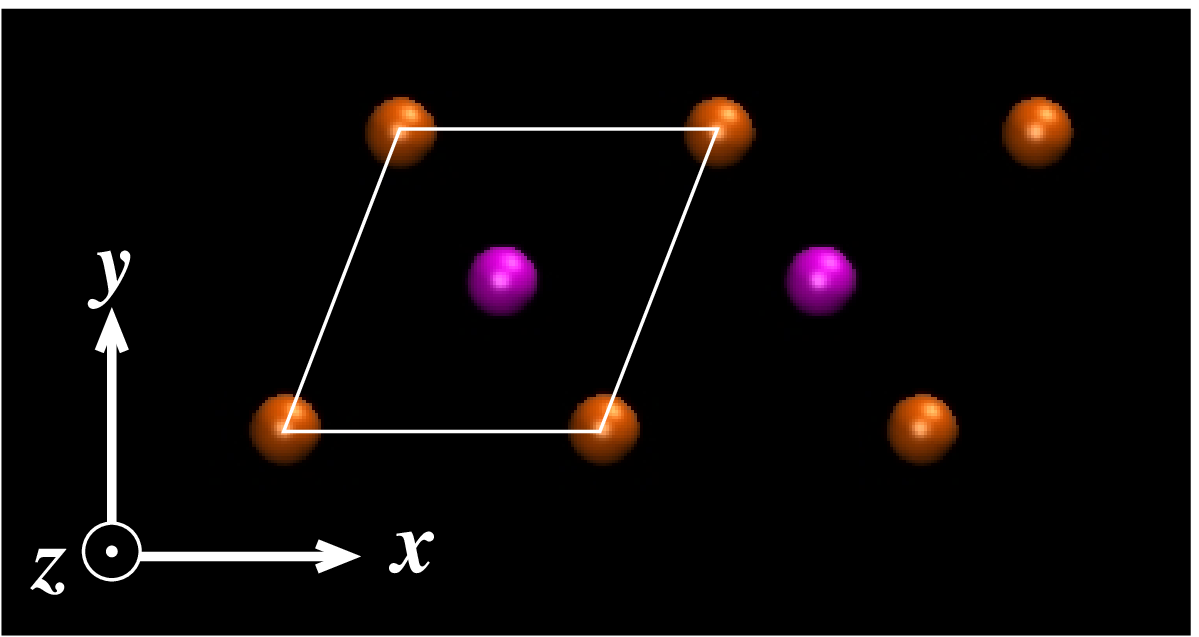} 
        & \includegraphics[width=3.1cm]{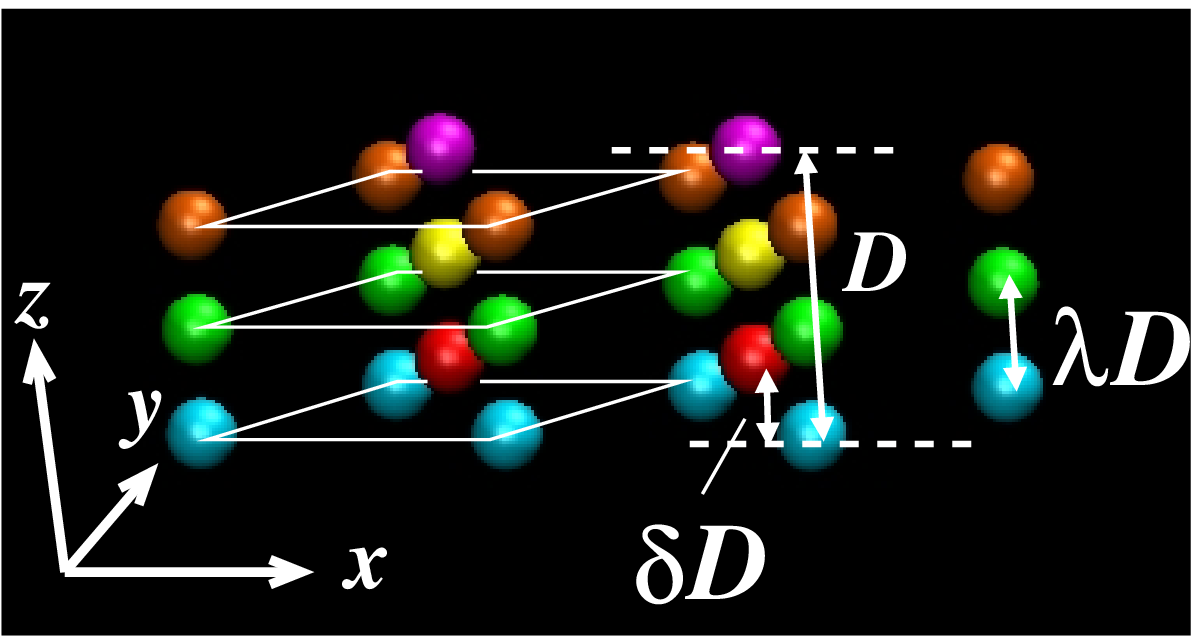} 
        \\\br
        
      \end{longtable}
    \end{indented}
  \end{center}
\end{landscape}
%

The primitive cells of all stable phases found in this work consist of one particle per layer. 
Each constitutive layer possesses the same basis shape ($\Delta$, $\square$ or $R$). 
These layers are shifted to each other, see table \ref{tab}.
Note that (for $m>3$) the layers become equidistant only in the limit $\eta \to \infty$.
A remarkable finding is the absence of prism phases (at $m=4$) that are encountered in
hard sphere systems \cite{Palberg_PRL_1997,Fortini} and Yukawa systems at finite 
screening \cite{Erdal_EPL}.

\begin{figure}[h!]
  \center
  \includegraphics[width=12cm]{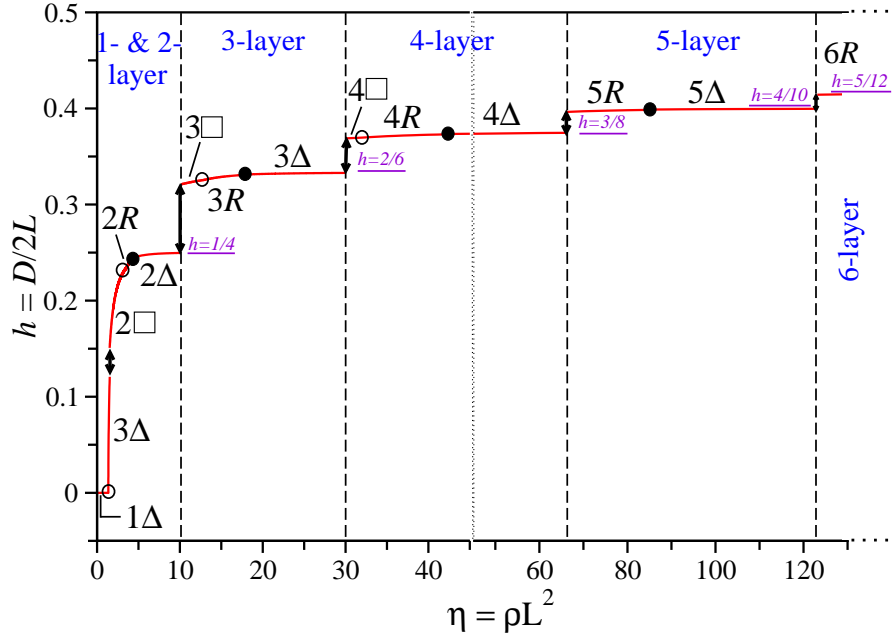}
  \caption{Order parameter $h$ of all stable crystalline phases. Empty circles denote a
    continuous transition, while the full circles mark a discontinuous one. The transitions 
    between different layer numbers, rendered as dashed lines, are also recorded as a first 
    order transition except $1\Delta \to 3\Delta$. 
    Apart of that, the underlined $h$-numbers give the limit $h$-value ($\eta \to \infty$), 
    for the case that no more phase transition to a higher layered structure occurs.
    The dotted line indicates a scale change in $\eta$-axis.
  }
  \label{fig7}
\end{figure}

A further overview of the full phase diagram ranging from triangular monolayer to 
rhombic hexalayer structures is shown in figure \ref{fig7} where the profile of 
$h(\eta)$ is also sketched. Empty circles indicate transitions of second order, 
while the full ones denote transitions of first order. In detail, for $3$- and $4$-layers, 
the transitions $3\square \to 3R$ and $4\square \to 4R$ occur continuously by 
continuously changing the angle $\theta$ between the two in plane basis vectors, 
in analogy to $2\square \to 2R$ (cf.\ figure \ref{fig4}), while all other transitions are 
discontinuous. Additionally, by the transitions $3R\to3\Delta$, $4R\to4\Delta$ and 
$5R\to5\Delta$, and by the transitions changing the layer number at 
$\eta=1.53$ ($3\Delta \to 2\square$), $\eta=10.14$ ($2\Delta \to 3\square$), 
$\eta=30.03$ ($3\Delta \to 4\square$), $\eta=66.24$ ($4\Delta \to 5R$) and $\eta=123.11$ 
($5\Delta \to 6R$) the distance between outermost layers exhibits a certain jump $\Delta h$ (indicated by 
thick arrows in figure \ref{fig7}). 
In fact, there is here no continuous transition present 
between two unequal layered phases as in the case of hard spheres 
\footnote{In the case of bilayered hard spheres, one can achieve a continuous layer increase 
  from $2\Delta$ to four-layered hcp-like and hcp(100) phase \cite{Manzano_2007,Fontecha_2007,Erdal_EPL}.}.

Furthermore, for high densities, the concrete lattice evolves to a continuous such that 
effects due to the concreteness get negligible. This means, electrostatically, 
that each layer of a $m$-layered structure is completely compensated by a certain 
part of background as much as $1/m$ of the whole.

In this paper we have dealt with a system consisting of particles ({\it macroions}) interacting 
via the unscreened Coulomb potential and of particles of opposite charge ({\it counterions}), which are 
homogeneously smeared out over a hard slit of width $L$, compensating the charge of the macroions. 
To determine the stability diagram of crystalline phases, we have performed lattice sum calculations 
of a set of candidates. As possible candidates we have taken into account 
phases with up to six layers ($m=1,\cdots,6$) whose primitive cell contains up to eight 
particles ($n=1, \cdots, 8$). Additionally, we considered the 
buckling phases from \cite{Chou_Nelson}, too. 
We have analyzed a regime up to $\eta \approx 130$ in our investigations. 
For small densities, we could trace the existence of the triangular monolayer $1\Delta$. 
Crossing a certain critical density $\eta_c$ the system buckles and evolves to a trilayered 
structure. This transition density is also calculated analytically by applying a Taylor expansion to 
the lattice sum for small separations. Furthermore the evolving of the layer separation from monolayer 
to trilayer could be characterized as $h(\eta) \sim (\eta -\eta_c)^{1/2}$, qualitatively.
Tuning the density upwards, we have noticed different stable bilayered structures, same as Wigner crystals. 
Beyond the bilayers, we could also find stable tri-, four-, five- and six-layers in square, 
rhombic and triangular bases. The final stability sequence for $m>4$ reads therefore: 
$mR \to m\Delta \to (m+1)R$ with a remarkable vanish of square-based phases, where the sequence 
for $m=3$ and $m=4$ is $m\square \to mR \to m\Delta \to (m+1)\square$.
While the stability domain of evenly layered phases gets larger 
with increasing $m$, the stability domain of square phases ($\square$) decreases for $m>2$ and 
disappears finally for $m>5$. On the other hand the stability domain of rhombic ($R$) and 
triangular ($\Delta$) phases increases both with growing $m>2$.

Apart of that, the transitions involved here are all of second order except $mR \to m\Delta$ and 
$m\Delta \to (m+1)\square$. The latter takes place discontinuously due to the order parameter 
$\theta$ and particle positions (as in the case of $nR \to n\Delta$) as well as 
with respect to $h$ (cf.\ \ref{fig7}).

\section{Conclusions}

To summarize: For slit-confined ions in a smeared background, we have determined
the ground state crystalline lattice as a function of the ion density up to
the six-layer regime. A complex cascade with buckled, squared and triangular
bi-, tri-, tetra-, penta- and hexalayers was found. The results are verifiable in
systems with classical ions in a background including charged colloids, dusty plasmas
and classical ions in a trap. One important conclusion is that the details
of multilayered structures depend crucially on the particle-background interaction. 
More future work is needed to include wall charges, wall particle attractions 
and effects of finite temperature \cite{Donko}. 
A detailed understanding of the stable crystalline structure as originating from the wall 
properties is desirable to construct filter devices \cite{Goedel} or 
optical band-gap crystals \cite{Pine}.

\ack
We thank T Palberg and S Apolinario for helpful discussions.
This work was supported by the DFG via the SFB TR6 (project D1).

 \appendix
 \section*{Appendix}
 \setcounter{section}{1}

 The total interaction energy per unit cell 
 of a crystalline unscreened Coulomb system can be written as 
 %
 %
 \begin{equation}
   \label{A1}
   U_C =  U_C^{s} + U_C^{c},
 \end{equation}
 %
 %
 where the unit cell consists of $n$ particles of charge $q$ located at ${\bf r}_i$.  
 The self energy $U_C^{s}$ in equation (\ref{A1}) stems from
 the interaction between a particle of the unit cell 
 and its own periodically repeated images. 
 The term $U_C^{s}$ in equation (\ref{A1}), is due to the  interaction
 between a particle of the unit cell and all other remaining $n-1$ particles of the cell 
 including their own images. 
 The convergence  involved in these sums is guaranteed 
 by the inclusion of a {\it surface} neutralizing background for each layer.
 Following the route of Br\'odka and Grzybowsky 
 (see equations $(16a)$, $(16b)$ and $(17)$ of reference \cite{Brodka}),
 $U_C^{s}$ and $U_C^{c}$ are given below. Therefore $U_C^{s}$ reads 
 %
 \begin{eqnarray}
   \fl  U_C^{s} = \frac{1}{|a_x|} n \frac{q^2}{\epsilon}
   \left\{ 4\left(\sum_{m,k = 1}^{\infty} \cos\left(2\pi k\frac{b_x}{a_x}m\right)
     K_0\left(2\pi k \left|\frac{b_y}{a_x}\right|m\right)\right)\right. 
   \nonumber \\
   \left. + \gamma_e -\ln\left(4\pi\left|\frac{a_x}{b_y}\right|\right) \right\} ,
   \label{A2}
 \end{eqnarray}
 %
 %
 with $\gamma_e = 0.577215665$ denoting the Euler-Mascheroni constant, 
 $K_0(x)$ the modified Bessel function of the second kind \cite{Abramowitz} and 
 $a_x$, $b_x$ and $b_y$ the corresponding $x$- and $y$-components 
 of the lattice vectors ${\bf a}$ and ${\bf b}$. Using the components 
 $x_{ij}=x_i - x_j$, $y_{ij}=y_i - y_j$ and $z_{ij}=z_i - z_j$ 
 of the relative separation vector ${\bf r}_{ij}$ 
 between cell particles $i$ and $j$, $U_C^{c}$ can be written as
 %
 %
 \begin{eqnarray}
   \fl U_C^{c} = \frac{1}{|a_x|}\sum_{i=1}^n
   \sum_{\begin{subarray}{h} j=1 \endline j>i \end{subarray}}^n \frac{q^2}{\epsilon}
   \nonumber \\
   \times \left\{ 4\sum_{m,k = 1}^{\infty}  
   \left[ \cos \left( 2\pi k\frac{x_{ij}+b_xm}{a_x} \right)\right.\right. 
   \nonumber \\
   \times K_0 \left( 2\pi k \left[ \frac{(y_{ij}+b_ym)^2+z_{ij}^2}{a_x^2} \right] ^{1/2} \right) 
   \nonumber \\
   + \cos \left( 2\pi k \frac{x_{ij}-b_xm}{a_x} \right) 
   \nonumber \\
   \left. \times K_0 \left( 2\pi k \left[\frac{(y_{ij}-b_ym)^2+z_{ij}^2}{a_x^2} \right] ^{1/2} \right) \right] 
   \nonumber \\
   + 4\sum_{k=1}^{\infty} \cos \left( 2\pi k \frac{x_{ij}}{a_x} \right)
   K_0 \left( 2\pi k \left[ \frac{y_{ij}^2+z_{ij}^2}{a_x^2} \right] ^{1/2} \right) 
   \nonumber \\
   \left. - \ln \left[ \cosh \left( 2\pi \left| \frac{z_{ij}}{b_y} \right| \right) 
     - \cos \left( 2\pi\frac{y_{ij}}{b_y} \right) \right] - \ln2 \right\}
   \label{A3}
 \end{eqnarray}
 %
 %
 for $(y_{ij},z_{ij}) \ne (0,0)$ and
 %
 %
 \begin{eqnarray}
   \fl  U_C^{c} = \frac{1}{|a_x|} \sum_{i=1}^n
   \sum_{\begin{subarray}{h} j=1 \endline j>i \end{subarray}}^n \frac{q^2}{\epsilon}
   \nonumber \\
   \times \left\{ 4\sum_{m=1}^{\infty} \sum_{k=1}^{\infty}\left[\cos\left(2\pi k\frac{x_{ij}+b_xm}{a_x}\right)
       K_0\left(2\pi k \left|\frac{b_ym}{a_x}\right|\right) \right.\right. 
   \nonumber \\
   \left.+ \cos\left(2\pi k \frac{x_{ij}-b_xm}{a_x} \right)
     K_0\left(2\pi k \left|\frac{b_ym}{a_x}\right|\right)\right]  
   \nonumber \\
   \left. -2\psi\left(\left|\frac{x_{ij}}{ax}\right|\right) -\pi\cot\left(\pi\left|\frac{x_{ij}}{a_x}\right|\right)
   -2\ln\left(4\pi\left|\frac{a_x}{b_y}\right|\right) \right\}
   \label{A4}
 \end{eqnarray}
 %
 %
 for $(y_{ij},z_{ij}) = (0,0)$, where $\psi(x)$ is the digamma function \cite{Abramowitz}.

 Being interested in the transition from mono- to trilayers, 
 we take as input the structure characteristics
 of the triangular phase $1\Delta$ into the lattice sums (\ref{A2})-(\ref{A4}): 
 $\theta = \pi/3$, $b_x/a_x = 0.5$, $b_y/a_x = \sqrt{3}/2$, $\gamma = 1$, $x_{12}/a_x = 0.5 = x_{23}/a_x$, 
 $y_{12}/b_y = 1/3 = y_{23}/b_y$, $x_{13}/a_x = 1$, $y_{13}/b_y = 2/3$, 
 $\rho = N/A = \frac{3}{a_x b_y} = \frac{2\sqrt{3}}{{a_x}^2}$ and therefore 
 ${a_x}^2 = \frac{2\sqrt{3}}{\rho} = \frac{2\sqrt{3} L^2}{\eta}$. 
 Here we consider for $1\Delta$ a multicell ($n=3$) consisting of three primitive cells, containing each 1 particle. 
 Thus, for a given $\eta$, the energy function $U_C$ depends now only on $z_{12}=hL=z_{23}$. 
 Taking this feature into account, the self energy and the cross energy finally read 
 %
 \begin{equation}
   U_C^{s} = \frac{1}{|a_x|} 3 \frac{q^2}{\epsilon}
   \left\{4 \sum_{m,k = 1}^{\infty} 
   \cos \left(\pi k  m \right) K_0\left(\pi k m\sqrt{3}\right) + \gamma_e -\ln\left(\frac{8\pi}{\sqrt{3}}\right) \right\} 
   \label{A5}
 \end{equation}
 %
 and
 \begin{eqnarray}
   \fl U_C^{c}(h) = \frac{1}{|a_x|}\sum_{i=1}^3
   \sum_{\begin{subarray}{h} j=1 \endline j>i \end{subarray}}^3 \frac{q^2}{\epsilon}
   \nonumber \\
   \times \left\{ 4\sum_{m,k = 1}^{\infty}  
     \left[ \cos \left(2\pi k \frac{x_{ij}+b_x m}{a_x} \right) K_0 \left( 2\pi k \left[{\lambda_{ij}^+}^2 
           + \beta_{ij}^2 h^2\right]^{1/2}\right) \right.\right.
   \nonumber \\
   \left. + \cos \left(2\pi k \frac{x_{ij}-b_x m}{a_x} \right) K_0 \left( 2\pi k \left[{\lambda_{ij}^-}^2 
         + \beta_{ij}^2 h^2\right]^{1/2} \right) \right] 
   \nonumber \\
   + 4\sum_{k=1}^{\infty} \cos \left(2 \pi k \frac{x_{ij}}{a_x} \right)
   K_0 \left(2\pi k \left[\frac{y_{ij}^2}{a_x^2} + \beta_{ij}^2 h^2\right]^{1/2}\right) 
   \nonumber \\
   \left.  - \ln \left[ \cosh \left( 2\pi |\phi_{ij}| h \right) - \cos \left( 2\pi\frac{y_{ij}}{b_y}\right) 
     \right] - \ln2 \right\} ,
   \label{A6}
 \end{eqnarray}
 where $\lambda_{12}^{\pm} = \sqrt{(y_{12} \pm b_y m)^2/a_x^2} = \sqrt{3/4}(1/3 \pm m) = \lambda_{23}^{\pm}$, 
 $\lambda_{13}^{\pm} = \sqrt{(y_{13} \pm b_y m)^2/a_x^2} = \sqrt{3/4}(2/3 \pm m)$, 
 $\beta_{12} = L/a_x = \beta_{23}$, $\beta_{13} = 2L/a_x$, $\phi_{12} = \sqrt{2\eta/3\sqrt{3}} = \phi_{23}$ 
 and $\phi_{13} = 2 \sqrt{2\eta/3\sqrt{3}}$.
 Before expanding the energy function at $h=0$, we first define 
 \begin{equation}
   \label{A7}
   f(h)^{\pm} = K_0 \left(2\pi k \left[{\lambda^{\pm}}^2 + \beta^2 h^2\right]^{1/2}\right) ,
 \end{equation}
 where the first four derivatives of $f(h)$ at $h=0$ are given as follows:
 \begin{eqnarray}
   \label{A8} 
   f(0)^{\pm} &= K_0 \left(2\pi k\lambda^{\pm}\right) ,
   \\
   \label{A9}
   f'(0)^{\pm} &= 0 ,
   \\
   \label{A10}
   f''(0)^{\pm} &= -K_{1}\left(2\pi k\lambda^{\pm}\right)\frac{2\pi k\beta^2}{\lambda^{\pm}} ,
   \\
   \label{A11}
   f'''(0)^{\pm} &= 0 ,
   \\
   \label{A12} 
   f''''(0)^{\pm} &= \left[K_0\left(2\pi k\lambda^{\pm}\right)2\pi k\lambda^{\pm} + 2K_1\left(2\pi k\lambda^{\pm}\right)\right]
   \frac{3\beta^4 2\pi k}{{\lambda^{\pm}}^3} .
 \end{eqnarray}
 Here, $K_1(x)$ is a modified Bessel function of the second kind \cite{Abramowitz}, too.
 Using a Taylor series and (\ref{A8})-(\ref{A12}), we now expand $U_C(h)$ from (\ref{A1})  
 at $h=0$ and achieve the final form of the energy:
 \begin{equation}
   \frac{\epsilon u(h)}{q^2 \sqrt \rho} = \underbrace{-1.960516 - 3.590668 \eta h^2 + 4.968827 
     \eta^2 h^4}_{\frac{U_C(h)\epsilon}{3q^{2}\sqrt{\rho}}}  + \frac{4}{3}\pi h^2 \sqrt{\eta}.
   \label{A13}
 \end{equation}
 The last term stems from (\ref{V_macroion_bg}), due to interactions with the background, respectively. 
 The coefficient $-1.960516$ corresponds to the static energy per particle of the triangular 
 lattice, which was already calculated in \cite{Bonsall}.


\end{document}